\documentclass[fleqn,usenatbib]{mnras}

\usepackage{newtxtext,newtxmath}

\usepackage[T1]{fontenc}

\DeclareRobustCommand{\VAN}[3]{#2}
\let\VANthebibliography\thebibliography
\def\thebibliography{\DeclareRobustCommand{\VAN}[3]{##3}\VANthebibliography}

\usepackage{graphicx}	
\usepackage{amsmath}	

\title[Off-lattice Monte Carlo]{Off-lattice Microscopic Monte Carlo Modeling of Molecular Hydrogen Formation on Carbonaceous Dust Grains}

\author[Satonkin et al.]{
N. A. Satonkin,$^{1}$\thanks{E-mail: n.a.satonkin@urfu.ru}
A. B. Ostrovskii,$^{1}$
A. A. Mozhegorov$^{1}$
A. F. Punanova$^{2}$\thanks{E-mail: punanovaanna@gmail.com}
and A. I. Vasyunin$^{1}$\thanks{anton.vasyunin@gmail.com}
\\
$^{1}$Research Laboratory for Astrochemistry, Ural Federal University, 48 Kuybysheva Street, Yekaterinburg, 620026, Russia\\
$^{2}$Onsala Space Observatory, Observatoriev\"agen 90, R\aa\"o, 43992 Onsala, Sweden
}

\date{Accepted XXX. Received YYY; in original form ZZZ}

\pubyear{\the\year{}}

\begin{document}
\label{firstpage}
\pagerange{\pageref{firstpage}--\pageref{lastpage}}
\maketitle

\begin{abstract}
In this work, we present an off-lattice Monte Carlo model of accretion and migration of hydrogen atoms on a rough surface of carbon dust grain. The migration of physisorbed atoms by means of thermal diffusion and quantum tunnelling through barriers between the surface potential minima is considered. The model is applied to simulations of molecular hydrogen formation in a cold interstellar medium for a temperature range 5---35~K. Eley---Rideal and Langmuir---Hinshelwood mechanisms for the formation of the H$_2$ molecule were taken into account. We found that the surface potential energy minima that hold the accreted hydrogen atoms (binding energy) has wide dispersion of its values. The minimum energy is three times smaller than the maximum energy for the uneven surface of the model grain. The large dispersion of the binding energies results in an extended range of temperatures where H$_2$ formation is efficient: 5---25~K. The dispersion of binding energies also reduces efficiency of diffusion due to tunnelling in comparison to that assumed in kinetic equation codes in which constant values of binding energies are employed. Thus, thermal hopping is the main source for the mobility of the hydrogen atoms in the presented off-lattice model. Finally, the model naturally provides the mean values for the ratio of binding-to-desorption energy. This ratio demonstrates weak dependence on temperature and is in the range of 0.5---0.6.
\end{abstract}




\section{Introduction}\label{sec:intro}

The formation of stars and planets is a fundamental problem in astrophysics. Although extensive studies done over several decades have already resulted in a general understanding of this process and shed light on many intricate details, there are still problems to be understood. It is accepted that the first step of star formation is the formation of giant clouds of molecular gas in the atomic interstellar medium. Since clouds mainly consist of molecular hydrogen (H$_{2}$) mixing with dust, understanding the formation process of this molecule is of great importance for studying the physics and chemistry of star and planet formation~\citep{1971ApJ...163..155H,1971ApJ...163..165H, Vidali13}. 

Several processes for the formation of molecular hydrogen in the interstellar medium were proposed~\citep{1963ApJ...138..393G,1963ApJ...138..408G,10.1046/j.1365-8711.1999.02480.x}. In principle, gas chemistry can lead to the formation of H$_{2}$~\citep{2003ApJ...584..331G,1983ApJ...271..632P}. In the early Universe, before the synthesis of heavy elements in the first stars, the formation of molecular hydrogen was only possible in gas-phase reactions between H atoms and positive or negative hydrogen ions (H$^{+}$, H$^{-}$). However, gas-phase formation of molecular hydrogen is relatively inefficient compared to other mechanisms that become available after the enrichment of the interstellar medium with heavy elements. Heavy elements can present in interstellar medium in various forms, including polycyclic aromatic hydrocarbons (PAHs) and interstellar dust grains~\citep{Allamandola_1999}.  The extraction of hydrogen from hydrogenated PAHs by atomic hydrogen accreted from the gas phase can be an efficient pathway for H$_{2}$ formation, particularly at higher temperatures. However, the formation of molecular hydrogen on the surface of interstellar dust grains is the most efficient and ubiquitous mechanism of H$_{2}$ formation in the interstellar medium.

Interstellar grains are omnipresent in the interstellar medium, with rare exceptions of some HII regions. Thorough analysis of dust extinction curves suggests that dust grains are of two main types~\citep{1984ApJ...285...89D}. Those are substantially silicate grains and carbonaceous grains~\citep{1977ApJ...217..425M,1989ApJ...341..808M,10.1093/mnras/211.3.719,10.1093/mnras/210.4.791,1980A&A....88..194H,1997A&A...323..566L,1988ESASP.281b.223A}. Grains are likely to form in atmospheres of giant cold stars and in dark molecular clouds\citep{1999ASIC..523..331T,2013A&A...558A..62J}. Specific conditions in dark clouds allow effective condensation and growth of dust grain nucleus. The growth of dust particles in dark molecular clouds can occur due to the adhesion of gas heavy species.

During the formation of molecular clouds, molecular hydrogen is formed on the bare surfaces of grains~\citep{WAKELAM20171,2004ApJ...611...40C,2004ASPC..309..529P}. However, in well-developed molecular clouds and especially in prestellar cores that represent the onset of star and planet formation, high gas density, low temperature of interstellar grains ($T<10$~K) and strong shielding of UV radiation ($A_{\mathrm{V}}>10$~mag) facilitate complex chemistry on grain surfaces~\citep{2009ARA&A..47..427H}. As a result, the blank grain surface quickly becomes covered with thick icy mantle that consists mainly from solid water, carbon monoxide and dioxide, methanol, ammonia and many less abundant species~\citep{Gibb_2000}. Icy mantles may include complex organic molecules, and thus they are one of the main objects of interest for research with the recently launched James Webb Space Telescope (JWST). However, complex chemistry in the icy mantles of interstellar grains is beyond the scope of this work. We focus on the H$_{2}$ formation process on bare grains that occurs during the formation of molecular clouds. Processes that lead to the formation of H$_{2}$ molecules from atomic hydrogen on rough blank grain surfaces with uneven binding sites for hydrogen atoms are objects of interest in this research. 

The goal of this study is to explore the impact of realistic spatial geometry of an interstellar dust grain that includes microscopic roughness of the grain surface on the chemical process of formation of molecular hydrogen that occurs on grain surfaces. For this purpose, we developed a variation of an off-lattice Monte Carlo method proposed in~\citep{2013ApJ...778..158G}. This approach represents a viable compromise between microscopic accuracy and computational efficiency. In the off-lattice Monte Carlo method, model particles that represent atoms or molecules are not fixed to a predefined lattice~\citep{2005A&A...434..599C,2005MNRAS.361..565C}. Instead, their mutual positions are determined by local minima of the surface potential. The surface potential is considered as the sum of the potentials of individual particles of ``grain surface'' in the vicinity of the particle of interest.

The paper is organized as follows. In Section~\ref{model}, the model setup is described in detail. Particular attention is paid to the assumptions about spatial structure of the model grain core and its interaction with gas phase species. In Section~\ref{results}, the results of the formation of molecular hydrogen are presented, including the discussion about the dominating mechanisms for the surface mobility of accreted hydrogen atoms. 

\section{Model description}\label{model}

\subsection{Interatomic interaction and construction of dust grain core}

The off-lattice Monte Carlo model presented in this study is based on the following principles. We assume that interacting species, species of the grain core, all are spheres. Interactions between all species in the model can be described using superposition of pair-wise Lennard-Jones potentials~\citep{1931PPS....43..461L}. In the current version of the model, there are two types of species. The first type are carbon atoms that are used to form static ``dust grain core''. The second type are hydrogen atoms that are allowed to accrete on grain surface from gas, desorb from the grain back to the gas, migrate on grain surface, and react with each other to form hydrogen molecules (H$_{2}$).

Classic expression commonly used for Lennard-Jones potential is as follows~\citep{ruthven1984principles}:
\begin{equation}
    U(r) = 4\cdot\epsilon\cdot \left[ \left(\frac{\sigma}{r}\right)^{12} - \left(\frac{\sigma}{r}\right)^{6} \right],
\end{equation}
where $r$ is a distance between two interacting particles, $\epsilon$ is a depth of potential well, and $\sigma$ is an effective size of the particle. Since in our model we are interested in the interaction of atoms of two chemical elements, namely, carbon atoms that form ''dust grain'' and hydrogen atoms that participate in chemical reaction leading to the formation of H$_{2}$ molecule, a ``mixture'' interaction potential shall also be defined~\cite{ruthven1984principles}:
\begin{equation}\label{mixture_potential}
    U(r) = 4\cdot\sqrt{\epsilon_{\mathrm{C}}\epsilon_{\mathrm{H}}}\cdot \left[ \left(\frac{\sigma_{\mathrm{C}}+\sigma_{\mathrm{H}}}{2r}\right)^{12} - \left(\frac{\sigma_{\mathrm{C}}+\sigma_{\mathrm{H}}}{2r}\right)^{6}\right].
\end{equation}
In the expression above, $r$ again is the distance between interacting atoms of hydrogen and carbon, $\sqrt{\epsilon_{\mathrm{C}}\epsilon_{\mathrm{H}}}$ and $(\sigma_{\mathrm{C}}+\sigma_{\mathrm{H}})/{2}$ are the parameters of the Lennard-Jones potential of interaction between carbon and hydrogen atoms. Based on van der Waals radii of hydrogen and carbon atoms equal to 1.2~\AA~and 1.7~\AA~correspondingly, $\sigma_{\mathrm{H}}$ = 2.14~\AA~ and $\sigma_{\mathrm{C}}$ = 3.03~\AA. The value of $\sqrt{\epsilon_{\mathrm{C}}\epsilon_{\mathrm{H}}}$ has been chosen to reproduce the average binding energy of hydrogen atoms on amorphous carbon equal to 660~K according to experimental work by~\citet[][]{Katz_1999}. The value takes into account the average number of pairwise interactions between hydrogen atom and surrounding carbon atoms located within the radius 2.5~$\times$~$\sigma_C$ around the hydrogen atom. Carbon atoms and hydrogen atoms are characterized by ($\epsilon_{\mathrm{C}}$, $\sigma_{\mathrm{C}}$), and ($\epsilon_{\mathrm{H}}$, $\sigma_{\mathrm{H}}$), respectively. In total, three types of pair interactions are possible in our model: C---C, H---H, and C---H. Thus, any atom residing near the grain surface, is affected by the total potential which is defined as a sum of pair-wise Lennard-Jones interaction potentials of this atom, with other atoms residing no further than $\sigma_{\mathrm{max}}$ from the atom of interest:
\begin{equation}\label{total_potential}
    U = \sum_{i=0}^{N}U_i(r_i).
\end{equation}
In the expression above, $U_i$ is a pair-wise interaction potential between the particle of interest and $i$th particle located no further than $\sigma_{max}$, $r_i$ is a distance between $i$th particle and the particle of interest.

The ``dust grain core'' is assembled from individual carbon atoms that interact with each other through the potentials described above. The atoms are added sequentially one-by-one to the initial ``seed'' that consist of a single carbon atom. Since our model operates in three dimensions, it is necessary also to  determine the spatial direction from which each added atom approaches the assembling grain core. The direction is chosen randomly such as trajectory of  approaching atom always intersect the centre of the spatial volume in the model. The centre coincides with the location of the ``seed'' carbon atom used to initiate the buildup of the grain core. The same approach to the choice of spatial directions is later used for accretion of hydrogen atoms from gas to the grain during the simulations of H$_{2}$ formation.

Each added carbon atom is allowed to relax in the collective potential of previously added atoms, i.e., to find the optimal spatial location with respect to other atoms. 
By optimal location, we assume a local minimum of the interaction potential between the accreting atom and carbon atoms forming the grain core and located within the radius of $\sigma_{\mathrm{max}}= 2.5\times\sigma_{\mathrm{C}}$ around the point of intersection between the trajectory of accreting atom and grain surface. $\sigma_{\mathrm{C}}$ stands for the size of a carbon atom. In principle, all atoms that form the grain core contribute to the interaction potential. However, since the attractive component of Lennard-Jones potential decreases as $r^{-6}$ with the distance between the interacting species, in practice only the core atoms located no farther than 2.5 of their sizes away of accreted hydrogen atom contribute appreciably to the resulting interaction potential. The same relaxation procedure is also applied to the hydrogen atoms during their accretion and migration on the surface of the dust grain core.

\subsection{Hydrogen chemistry on grain surface}

Let us now consider in details treatment of processes that involve hydrogen atoms included in the model. Since no gas-phase chemistry is included into the model, flux of atomic hydrogen from gas to grain is constant during simulations. The accretions rate of atomic hydrogen from the gas phase is calculated using the following expression:
\begin{equation}\label{kacc}
    k_{\mathrm{acc}} = n\cdot\Sigma\cdot v\cdot S(T,\;T_{\mathrm{d}}),
\end{equation}
where $n$ is gas density in cm$^{-3}$, $\Sigma$ is mean cross-section of the grain, $v$ is mean velocity of hydrogen atoms, and $S(T,\;T_{\mathrm{d}})$ is sticking probability. $S(T,\;T_{\mathrm{d}})$ is calculated using the Expression (3.7) in~\cite{1979ApJS...41..555H}.

\begin{figure*}
\begin{tabular}{ccc}
    \includegraphics[height=3.5cm,keepaspectratio]{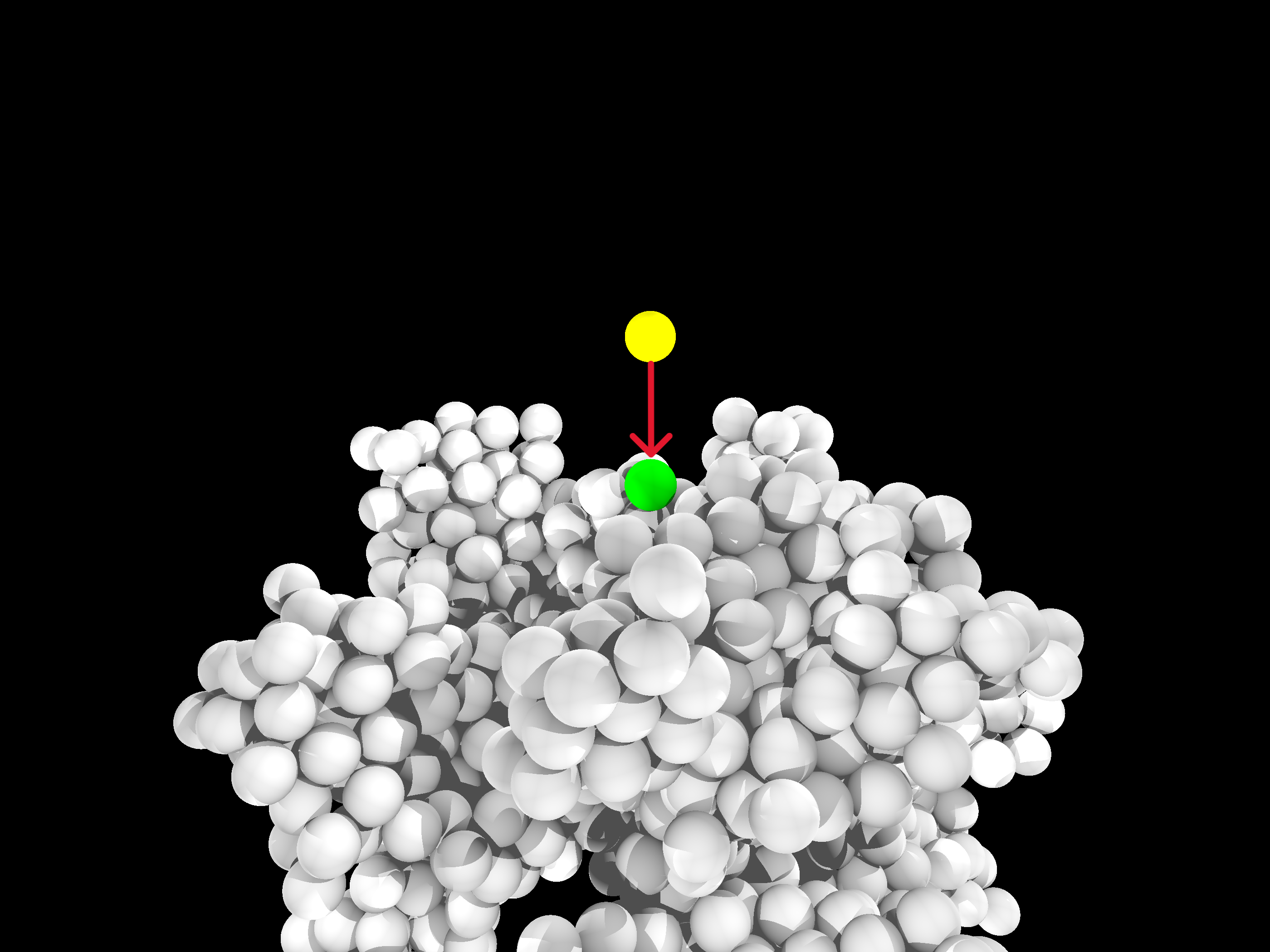} & \includegraphics[height=3.5cm,keepaspectratio]{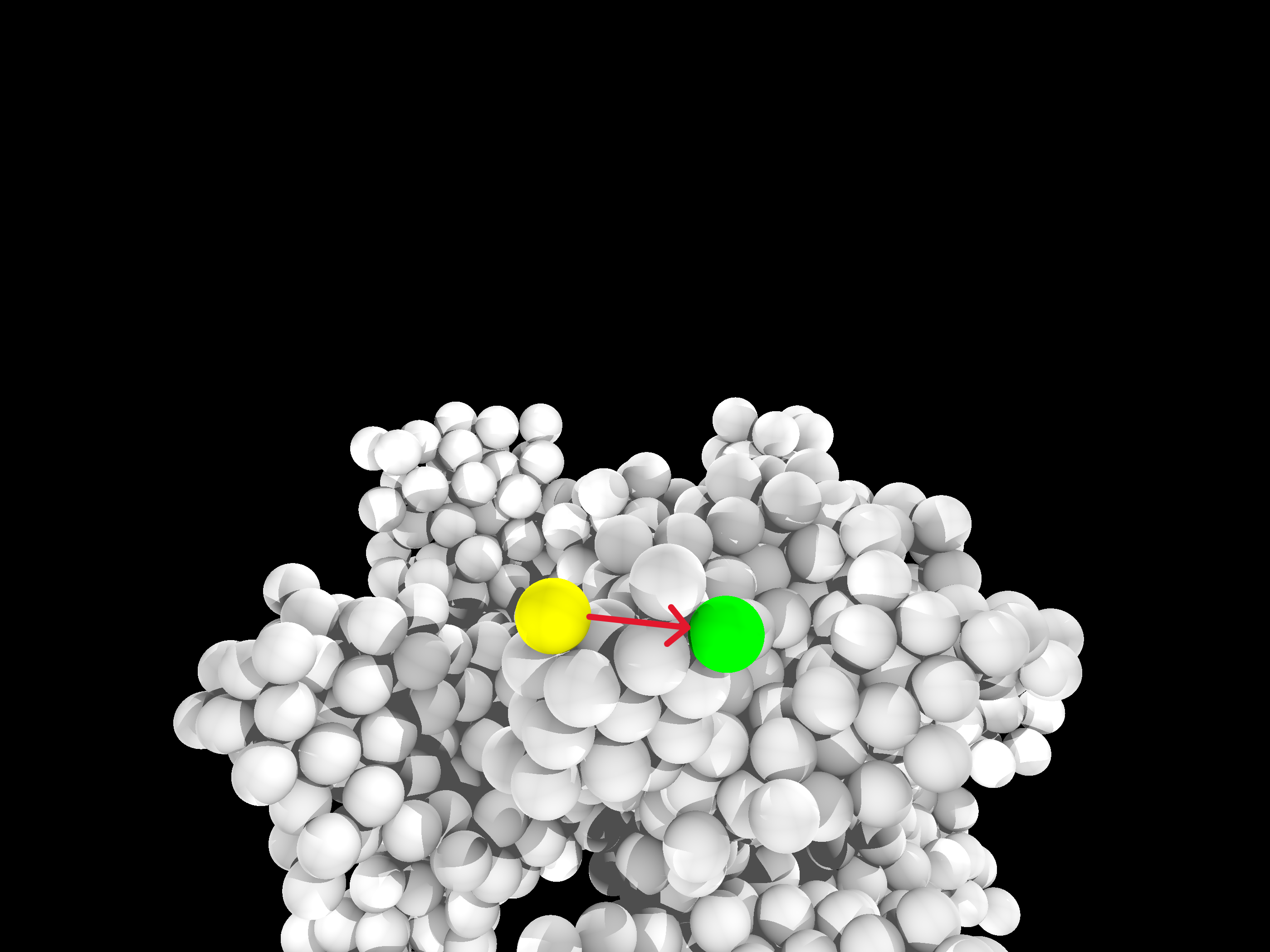} & \includegraphics[height=3.5cm,keepaspectratio]{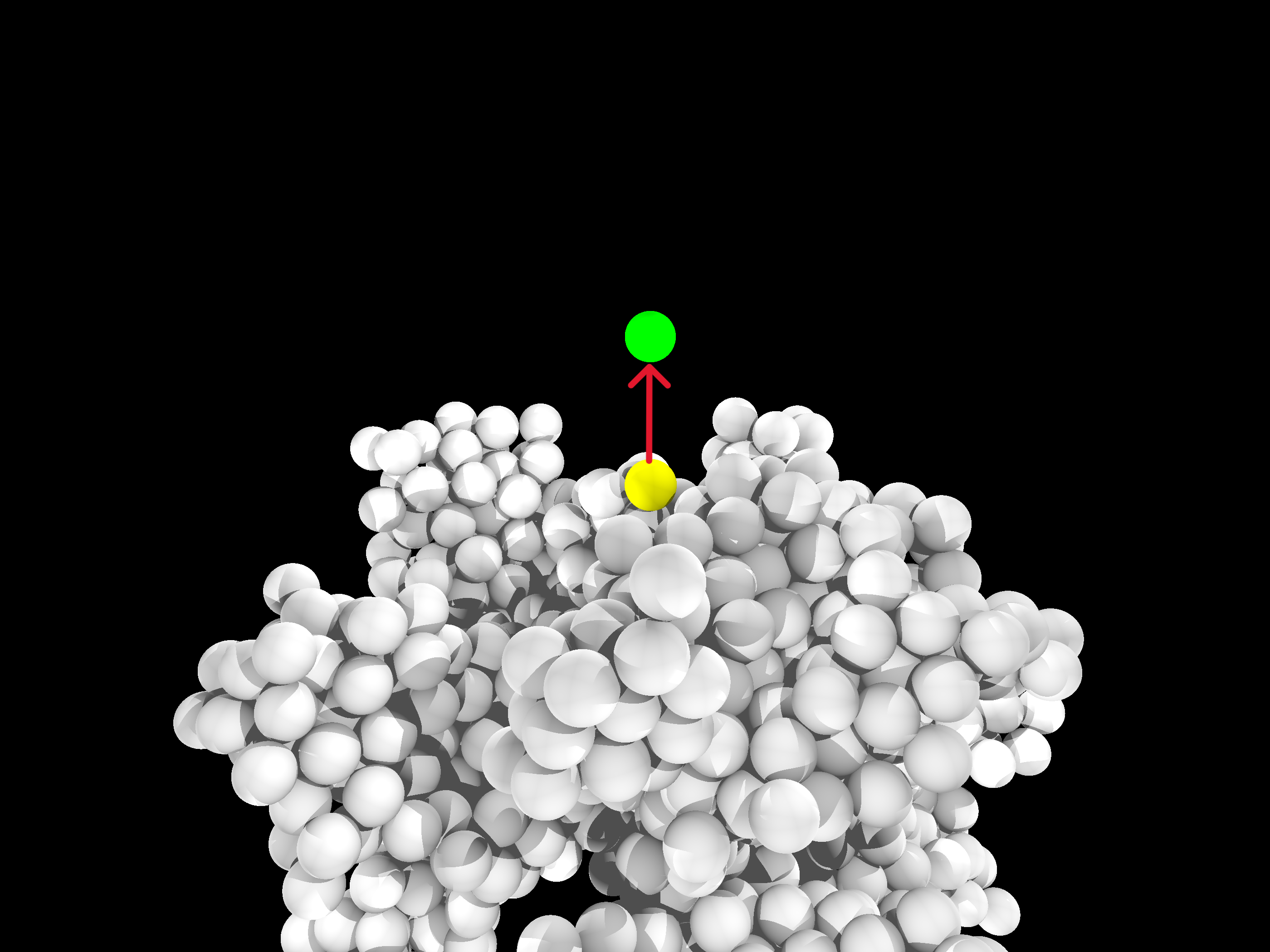} \\
    (a) & (b) & (c) \\
\end{tabular}
\caption{Grain-related processes considered in the model: accretion (a), diffusion (b) and desorption (c) of hydrogen atoms. Diffusion is possible due to thermal hopping and quantum migration. Only thermal desorption is considered.
\label{fig:h2_processes}}
\end{figure*}

After accretion, several types of events may happen to any accreted hydrogen atom: it may desorb back to the gas phase, it may ``hop'' thermally to the adjacent potential minimum on a grain surface, or it can move to the adjacent potential well via quantum tunnelling through the potential barrier between currently occupied potential well and the adjacent one (see Figure~\ref{fig:h2_processes}). The search for potential minima that could be possible destinations for migration of hydrogen atom is performed within the distance of 2.5$\times\sigma_{max}$ from the current location of hydrogen atom. All found potential minima are considered as possible destinations for migration via thermal hopping. However, only those potential minima that equal to or deeper than potential minimum currently occupied by the hydrogen atom of interest are considered as possible destinations for migration via quantum tunnelling. Note that such determination of the direction of migration differs from the approach proposed in~\citet[][]{2013ApJ...778..158G}. There, migration is only possible through the saddle points of surface potential to adjacent potential minima thus making their model not fully lattice-free.
The rates for tunnelling and thermal hopping for each found potential minimum, as well as desorption for an accreted hydrogen atom are calculated using the following expressions~\citep{1992ApJS...82..167H}:
\begin{equation}\label{ktunn}
    k_{\mathrm{tunn}} = \nu\cdot \exp\left(-\frac{2a}{\hbar}\,\sqrt{2\cdot m\cdot E_{\mathrm{bind}}}\right),
\end{equation}

\begin{equation}\label{khop}
    k_{\mathrm{hop}} = \nu\cdot \exp\left(-\frac{E_{\mathrm{hop}}}{T}\right),
\end{equation}

\begin{equation}\label{kdes}
    k_{\mathrm{des}} = \nu\cdot \exp\left(-\frac{E_{\mathrm{bind}}}{T}\right).
\end{equation}
In the expressions above, $\nu$ is a characteristic vibration frequency for hydrogen atoms, $m$ is a mass of hydrogen atom, $a$ is width of a potential barrier (i.e., distance) between potential minima (binding sites), T is grain temperature, E$_{bind}$ is a binding energy of a hydrogen atom, and E$_{hop}$ is a energy barrier against thermal hopping.

It is important to note that in our off-lattice model there are no predefined values of binding and diffusion energies for hydrogen atoms. Therefore, there is also no predefined binding-to-diffusion energy ratio. Instead, binding energies are taken individually for each accreted hydrogen atom as values of total interaction potential (\ref{total_potential}) between the atom and currently occupied location on surface. Energy barriers for thermal hopping in turn are calculated as follows. As described above, the total potential of interaction between the hydrogen atom and (\ref{total_potential}) is a sum of pair-wise interaction potentials between the hydrogen atom and adjacent carbon atoms. The potential migration destination sites are also surrounded by carbon atoms. Some of them are the same as those surrounding the current location of the hydrogen atom, while the others are different. Thus, in case of hopping from the currently occupied site to the new site, some of $C - H$ interatomic bonds will be broken (i.e., the value of corresponding mixture potentials ((\ref{mixture_potential}) will become near-zero), while some other bonds will be weakened (i.e., the value of (\ref{mixture_potential}) will become smaller). Therefore, we define the value of energy barrier for thermal hopping E$_{hop}$ as a sum of broken bonds plus the sum of differences between the new values of weakened bonds and their initial values.

Temporal evolution in the considered off-lattice Monte Carlo model is simulated using the Gillespie's stochastic simulation algorithm~\cite{GILLESPIE1976403} in its ``first-reaction method'' formulation. The simulation procedure consists of the following steps.
\begin{itemize}
    \item The rate of accretion of hydrogen atoms from the gas phase $k_{acc}$ is calculated
    \item For each accreted hydrogen atom, rates of desorption $k_{des}$, as well as rates of tunnelling $k_{tunn}$ and $k_{hop}$ thermal hopping to all achievable diffusion destinations are calculated
    \item ``Waiting times'' for each possible event are calculated following~\cite{2007A&A...469..973C}:
    \begin{equation}
        \tau_{\mathrm{acc}} = -\frac{\ln(X_{1})}{k_{\mathrm{acc}}}, 
    \end{equation}
    \begin{equation}
        \tau_{\mathrm{des}} = -\frac{\ln(X_{2})}{k_{\mathrm{des}}}, 
    \end{equation}
    \begin{equation}
        \tau_{\mathrm{tunn}} = -\frac{\ln(X_{3})}{k_{\mathrm{tunn}}},
    \end{equation}
    \begin{equation}
        \tau_{\mathrm{hop}} = -\frac{\ln(X_{4})}{k_{\mathrm{hop}}}. 
    \end{equation}
    where $X_{1,2,3,4}$ are the uniformly distributed random numbers in the range (0...1). Note that waiting times for tunnelling and for thermal hopping are calculated independently. Waiting times for desorption are only calculated for those accreted hydrogen atoms that are not covered by carbon atoms. Such situation can rarely happen as a result of diffusion of hydrogen atoms on the surface of a grain core of complex shape.
    \item The process with minimum waiting time $\tau$ is selected to happen on a current time step. Depending on the process, the following is performed.
    \begin{itemize}
        \item If the accretion event is chosen, new hydrogen atom arrived from a random direction is ``landed'' on grain surface to the optimal location as described above. For any hydrogen atoms residing within 2.5$\times\sigma_{max}$ from the landed atom, potential migration destinations are recalculated. 
        \item If the desorption event is chosen, the atom is eliminated from the model. The impact of its potential on surrounding atoms is excluded. 
        \item If the migration of hydrogen atom is chosen due to quantum tunnelling or due to thermal hopping, the diffusing atom is moved to the new potential minimum. For all atoms residing within the 2.5$\times\sigma_{\mathrm{max}}$ from the landed atom, total potentials $U$ are recalculated, and new possible destinations their for migration are determined. Note that migration due to tunnelling is only possible to those destinations that are characterized by the same or lower value of total potential $U$. On the contrary, migration due to thermal hopping is possible to any reachable potential minimum.
    \end{itemize}
    \item After the chosen event processed, the clock in the model is increased by the corresponding $\tau$, and algorithm returns to the choice of next event.
    \item Any approach on grain surface of two hydrogen atoms within 2.4~\AA~ of each other either as a result of diffusion or accretion of the second atom from the gas phase, is considered as formation of H$_2$ molecule. The first situation corresponds to the H$_2$ formation via the Langmuir-Hinshelwood mechanism, while the second corresponds to the Eley-Rideal mechanism. The corresponding two hydrogen atoms are removed from simulation, as well as the impact of their potentials on adjacent atoms. This mimics immediate desorption of H$_2$ molecules upon formation. The assumption of immediate desorption of newly formed H$_{2}$ molecule is justified by the fact that the reaction ${\rm H}~+~{\rm H}~\rightarrow~{\rm H_2}$ is exothermic with energy yield of 4.6 eV. This is an equivalent of 53000~K which is order of magnitude higher than the deepest potential minima in the model. At least one third of this energy is retained by the newly formed molecule~\citep[][]{Pantaleone_ea21} ensuring its desorption. Therefore, in the current version of the model we do not simulate the formation of hydrogen molecule itself. Accurate simulation of the assembly of H$_2$ molecule goes beyond our simple model, and is not required within the scope of this work.
\end{itemize}
As it can be inferred from the description above, the off-lattice simulation algorithm is computationally intensive. To calculate one thousand years of chemical evolution approximately one week of CPU time on Intel Xeon Gold 6154 is needed. Since the current version of the developed off-lattice model is not capable of building up an icy mantle on grain core, it was applied to exploration of the formation of molecular hydrogen under conditions in the diffuse interstellar medium. We assumed that the bare dust grain core is surrounded by mostly atomic gas with number density of hydrogen nuclei of 10~cm$^{-3}$. The formation of molecular hydrogen on a grain was explored in a range of temperatures [5---35]~K. Note that fluctuations of temperature due to stochastic heating of grains by UV photons are ignored in the model. Although the lowest grain temperatures are surely too low for diffuse medium, somewhat extended range of temperatures allowed us to better explore some important trends in model behaviour. Visual extinction and strength of the UV field are not explicitly present in the model, since no photoprocesses are currently included. However, for such a low-density gas the typical values of $A_{\mathrm{V}}$ must be below unity: $A_{\mathrm{V}}<1$~mag.
\begin{figure*}
\begin{tabular}{cc}
    \includegraphics[height=6cm,keepaspectratio]{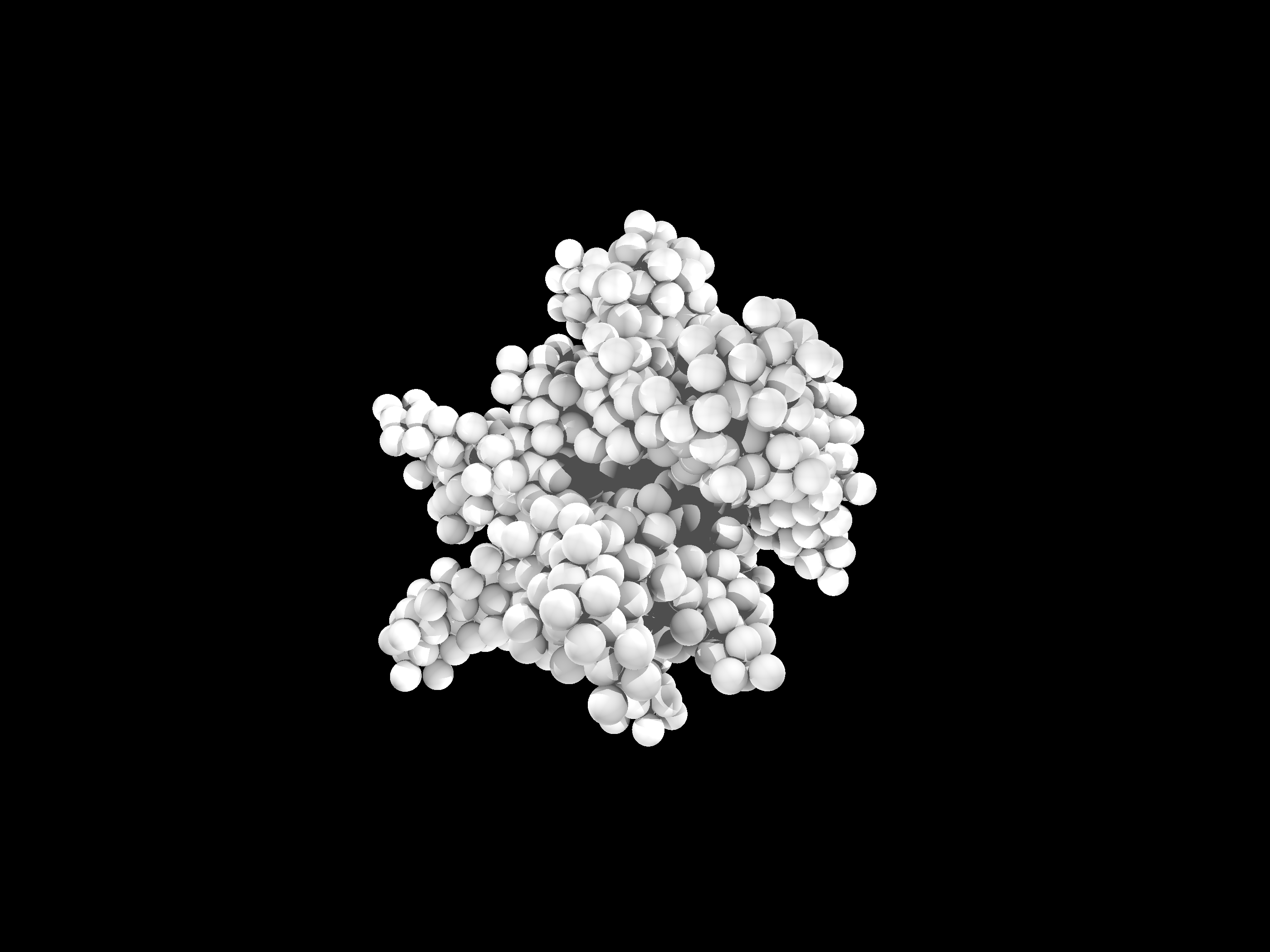} & \includegraphics[height=6cm,keepaspectratio]{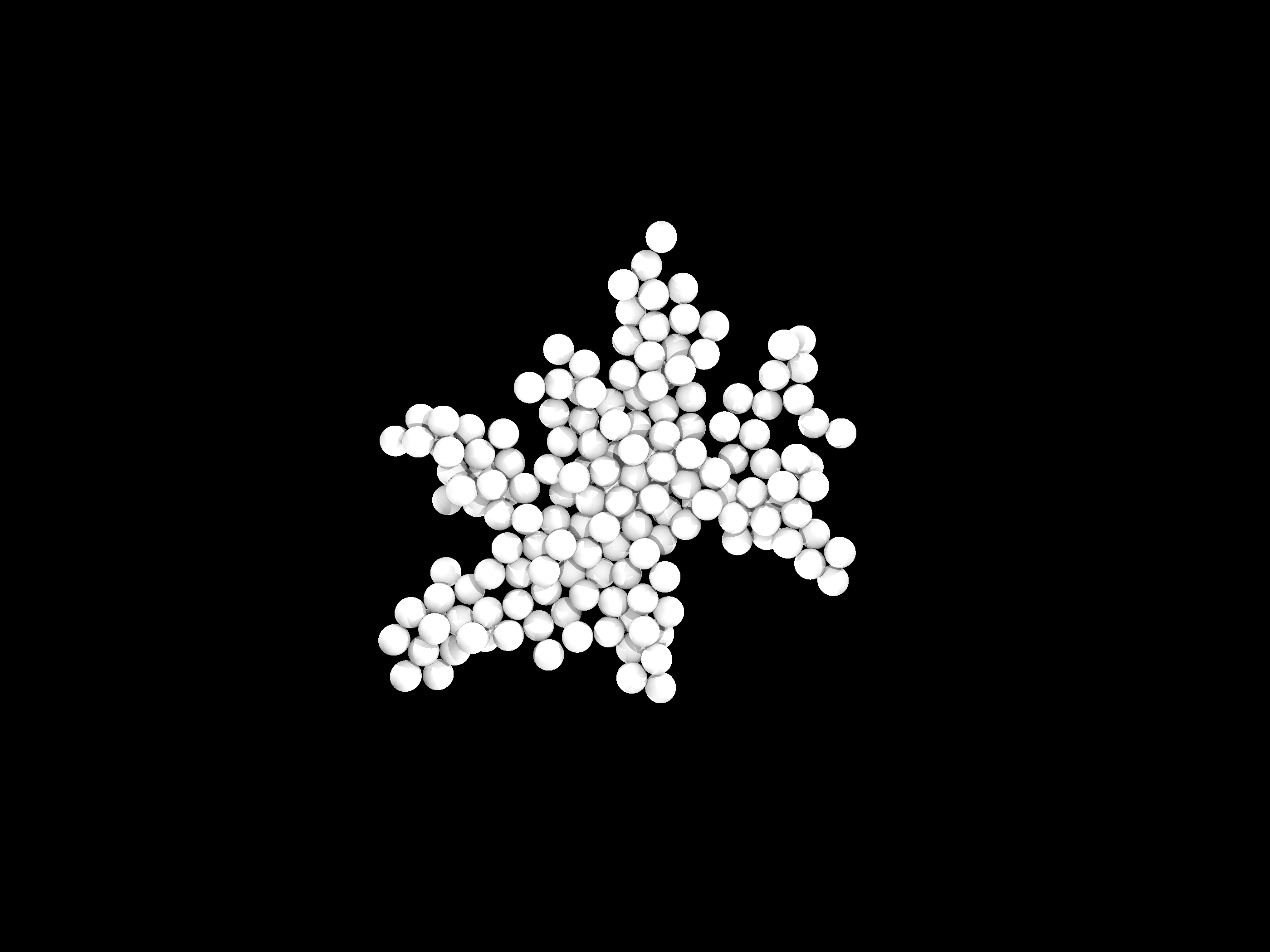} \\
    (a) & (b) \\
    \includegraphics[height=6cm,keepaspectratio]{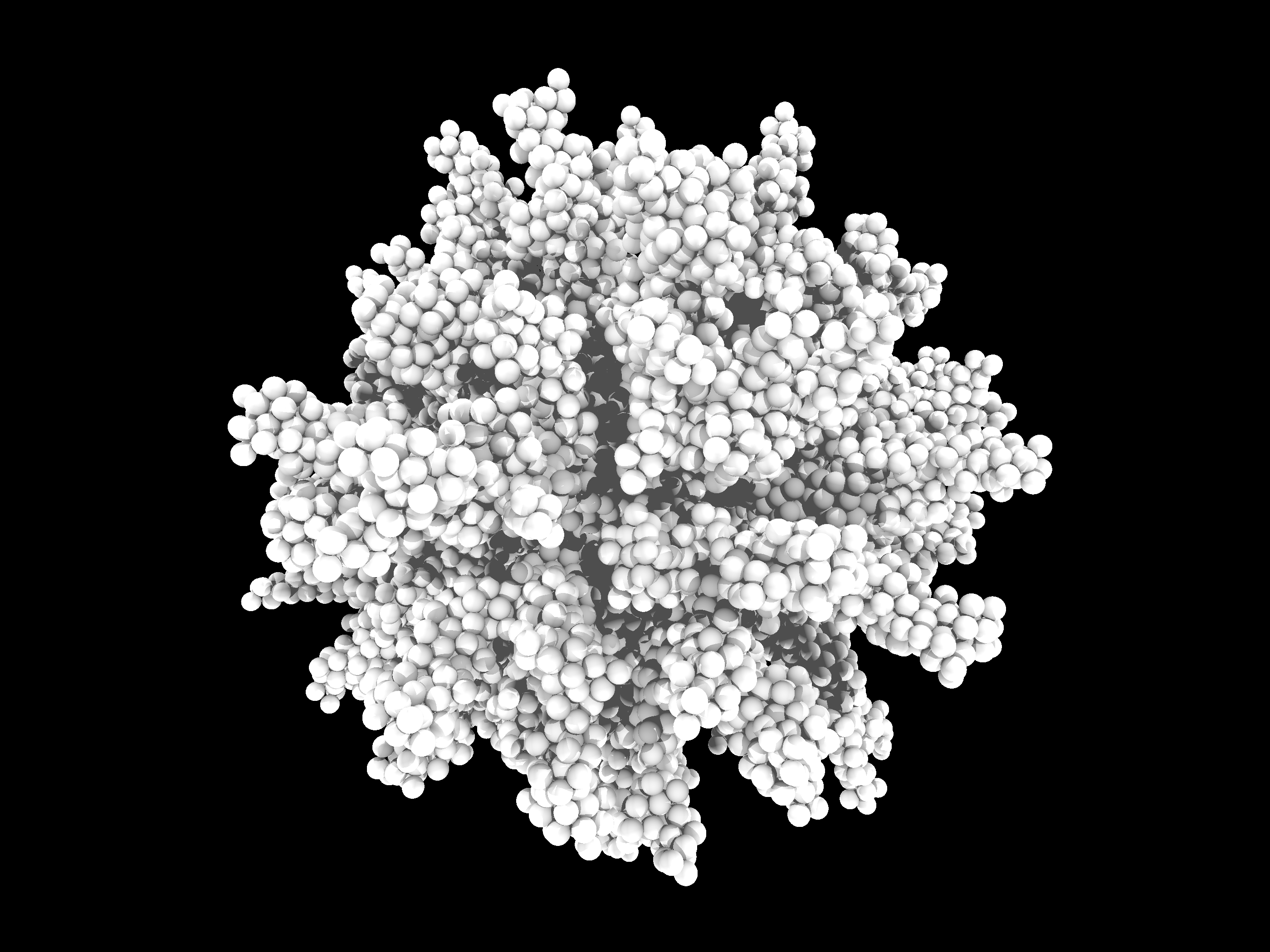} & \includegraphics[height=6cm,keepaspectratio]{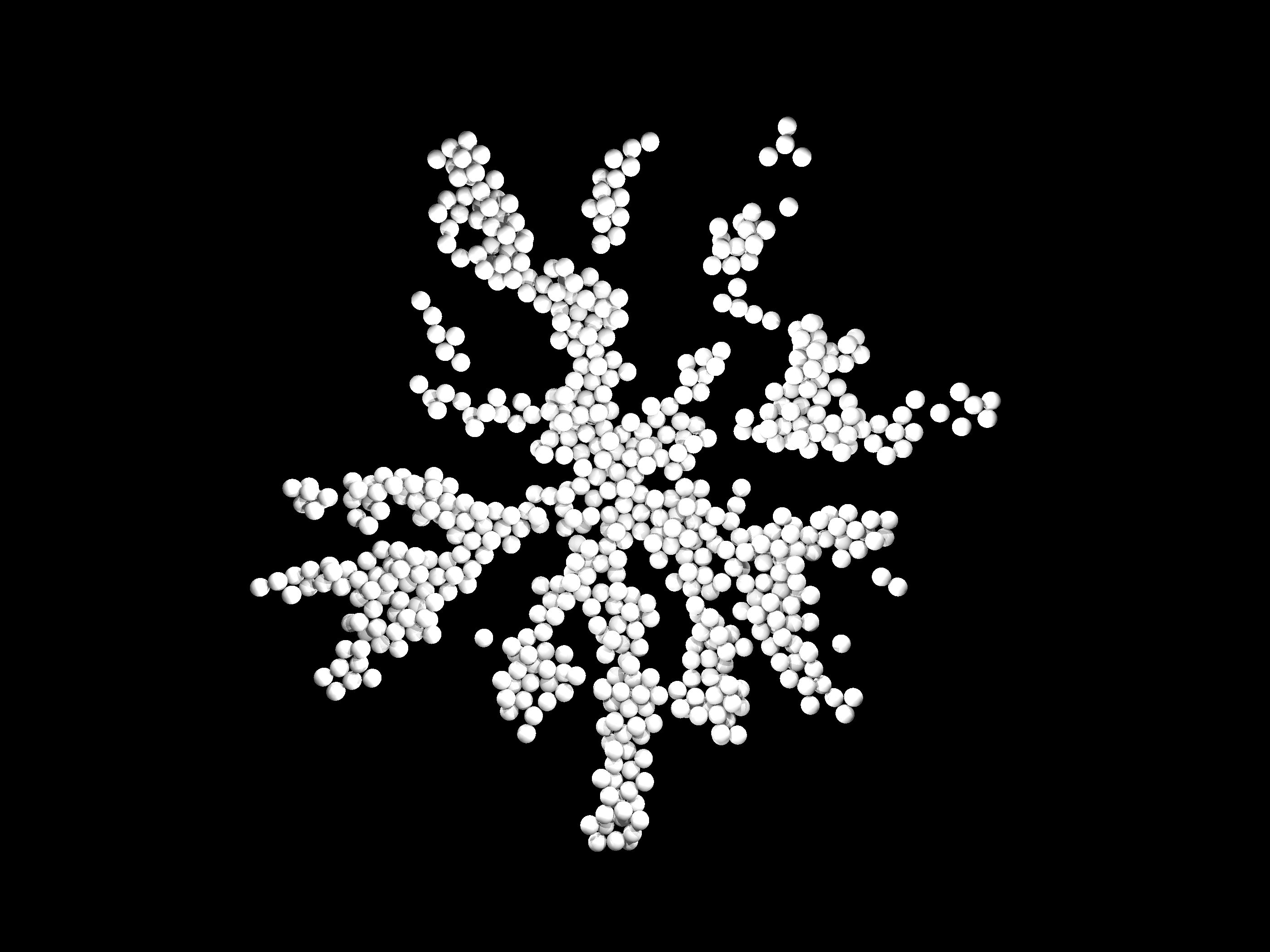} \\
    (c) & (d) \\ 
\end{tabular}
\caption{(a) ``Small'' grain core assembled up of 1000 carbon atoms; (b) cross-section of a ``small'' grain core; (c) 'large' grain core assembled up of 10000 carbon atoms; (d) cross-section of a ``large'' grain core.
\label{fig:grain_cores}}
\end{figure*}

\begin{figure}
    \centering
    \includegraphics[height=5.5cm,keepaspectratio]{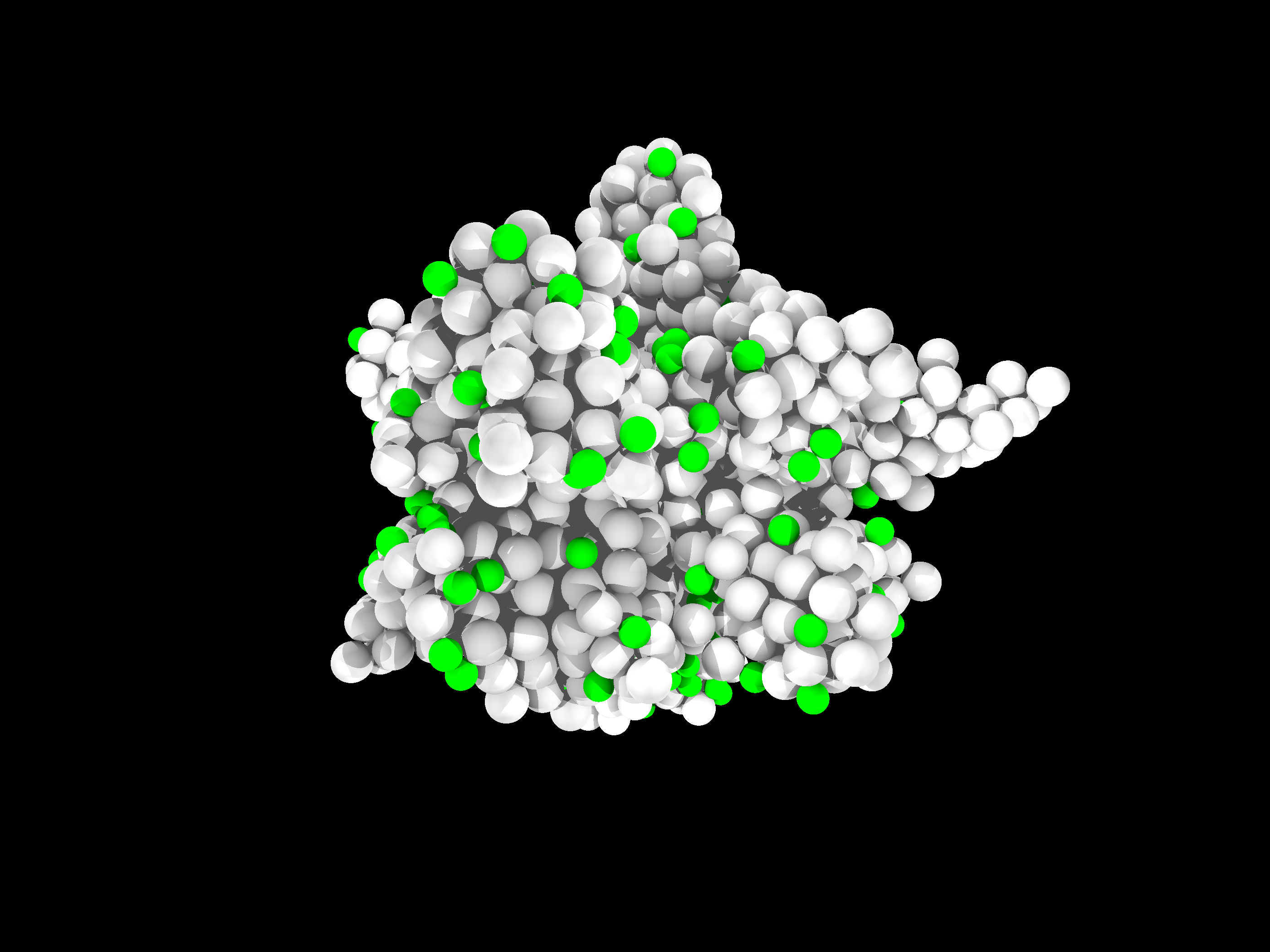}
    \caption{An example of a $6\times$10$^{-7}$~cm grain core with temperature of 5~K covered with hydrogen atoms. On average, $\sim$130 hydrogen atoms residing simultaneously on grain surface.}
    \label{fig:grain_with_H_atoms_5K}
\end{figure}

\section{Results}\label{results}

Let us now consider the results obtained with the model described above. In Figure~\ref{fig:grain_cores}, examples of amorphous grain cores that consist of carbon atoms and constructed using the off-lattice model described above are presented. In the top left panel, a ``small core'' that consists of 1000 carbon atoms is shown, while in the bottom left panel a ``large core'' that includes 10~000 carbon atoms is presented. In the right column of the figure, corresponding cross-sections of the cores are shown. It can be seen that generated grains have highly irregular shape and curvy surface. The latter is the reason for variations of binding energies for hydrogen atoms at different surface sites. Note also that the sizes of generated grains are one--two orders of magnitude smaller than median size of grains in MRN distribution. The latter is 0.1~$\mu$m, i.e., 100~nm. The size of a ``small'' generated grain is $\sim$6~nm, while the size of a ``large'' grain is $\sim$14~nm. However, the grains of different sizes are likely self-similar. Thus, the results obtained in this work can be safely generalized to larger grains. We tested this by comparing the simulation results for our ``small'' and ``large'' grains. The results appeared to be similar within the simulation noise. Thus, in the following we only present simulation results for the ``small'' grain.
\begin{figure}
\includegraphics[height=5.5cm,keepaspectratio]{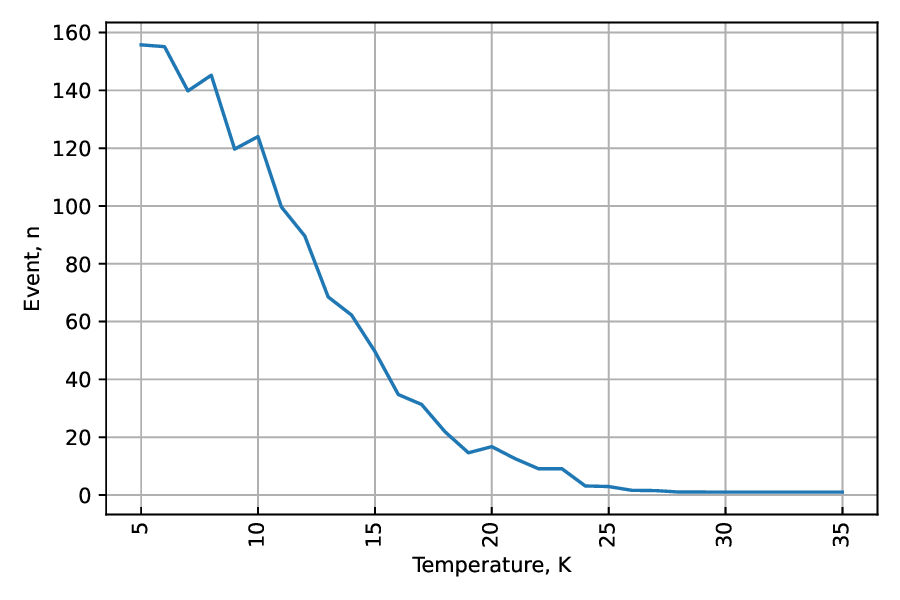}
\caption{Average number of atoms simultaneously residing on grain during the simulations vs. temperature.\label{fig:Natoms_vs_T}}
\end{figure}
In the course of simulations, hydrogen atoms accrete on grain core, diffuse of it, recombine to H$_2$ molecules or desorb back to the gas. At low temperatures, a significant number of hydrogen atoms reside on surface simultaneously, thus allowing Eley-Rideal mechanism of H$_2$ formation possible. In Figure~\ref{fig:grain_with_H_atoms_5K}, a snapshot of a small grain core with temperature of 5~K covered with hydrogen atoms is shown. On average, about 130 atoms residing on its surface simultaneously. At higher temperatures, the surface coverage by H atoms drops quickly (see Figure~\ref{fig:Natoms_vs_T}), and only diffusive mechanisms of H$_2$ formation have non-zero efficiencies.

\begin{figure}
\includegraphics[height=5.5cm,keepaspectratio]{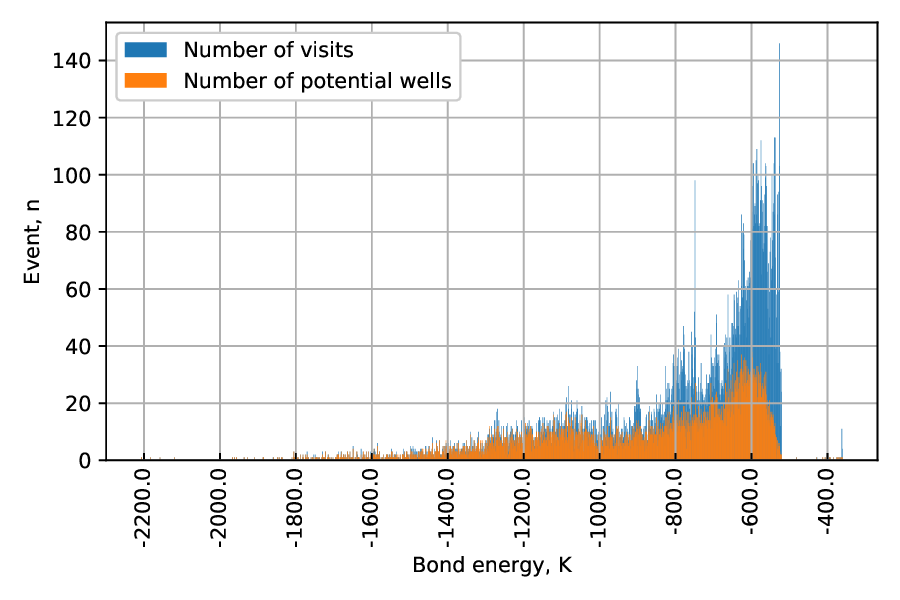}
\caption{Depth distribution of visited potential wells on the ``large'' amorphous grain core at 5~K. Orange - number of potential wells with given depth; blue - number of visits for each potential well with given energy. \label{fig:distribution_of_potential_wells}}
\end{figure}
During the simulation, statistical information about every event happened to each hydrogen atom on grain has been collected. This information includes depths of visited surface potential minima, heights of barriers against thermal diffusion, widths of potential barriers passed through during diffusion via quantum tunnelling. In Figure~\ref{fig:distribution_of_potential_wells}, the distribution of values of potential minima on grain surface visited by hydrogen atoms during the simulations of H$_2$ formation at 5~K is shown. It can be seen that potential minima, i.e., the binding energies of hydrogen atoms on an uneven surface have different values continuously distributed in a range of approx. 520~K -- 1\,800~K. This is in contrast to a widely utilized assumption that adsorption of a certain chemical species on grain surface can be characterized by a single value of binding energy as first proposed in~\cite{1992ApJS...82..167H} and then implemented in a number of rate equations-based astrochemical models, but in agreement with a number of accurate quantum-chemical simulations that also predict distribution of binding energies for astrochemically important species on e.g. amorphous or crystalline water ice~\citep[][]{Bovolenta_ea20, Ferrero_ea2020, Duflot_ea21, Bovolenta_ea22, Germain_ea22, Tinacci_ea22}. Still, the absolute majority of rate equations-based models of interstellar chemistry includes single values of binding energies of a given species on a given type of surface. Several exceptions exist, that clearly demonstrate that including distribution of binding energies in astrochemical modeling leads to important results such as more distant from the central star location of the water snowline it protoplanetary disks~\citep[][]{Tinacci_ea23} or possible reason for lacking detections of H$_2$S molecule in interstellar ices~\citep[][]{Bariosco_ea24}. In more detailed microscopic on-lattice Monte Carlo models of grain chemistry~\cite{2005MNRAS.361..565C}, binding energies of species depend on local geometry of surface, and can differ for same chemical species in different locations on surface. However, in on-lattice models the number of possible local geometries of surface is finite, so the distribution of binding energies is discrete.

In the Figure~\ref{fig:distribution_of_potential_wells}, besides the distribution of potential minima by depth, the numbers of visits of minima of every depth by hydrogen atoms are shown. It can be seen that weak potential minima tend to be visited more frequently than strong minima. This is explained by the fact that at low grain temperature, a weak binding site can be occupied, left and occupied again by the same hydrogen atom multiple times. In contrast, strong binding sites tend to be occupied more permanently by hydrogen atoms: higher potential barriers that surround such sites make diffusion or desorption of an atom stuck in the site quite improbable. Atoms in such deep potential wells however do not reside there permanently. Eventually, they are approached either by a diffusing hydrogen atom or hydrogen atom that is accreting from a ``lucky'' direction. Either cases result in the formation of H$_2$ molecule and vacating the strong binding site. The existing of strong binding sites increase the number of hydrogen atoms simultaneously residing on surface, as well as span of temperatures where the H$_2$ formation via the surface H~+~H~$\rightarrow$~H$_2$ reaction has non-zero efficiency.

Thermal hopping of hydrogen atoms on surface is controlled by the energies of potential barriers against diffusion. In macroscopic rate equations-based models, the height of barriers against hopping is normally a certain fraction of species's binding energy: $E_{\mathrm{hop}}=X\times E_{\mathrm{bind}}$. Here, $X$ is a poorly known parameter. Its value is typically taken in the range [0.2---0.8]~\citep[see][for review]{2022ESC.....6..597M}. In contrast, in our microscopic off-lattice model, the hopping energies are individual for every possible direction of hopping out of every binding site. Therefore, similarly to binding energies, values of hopping energies also distributed in a certain range.
\begin{figure*}
\begin{tabular}{ccc}
    \includegraphics[height=3.5cm,keepaspectratio]{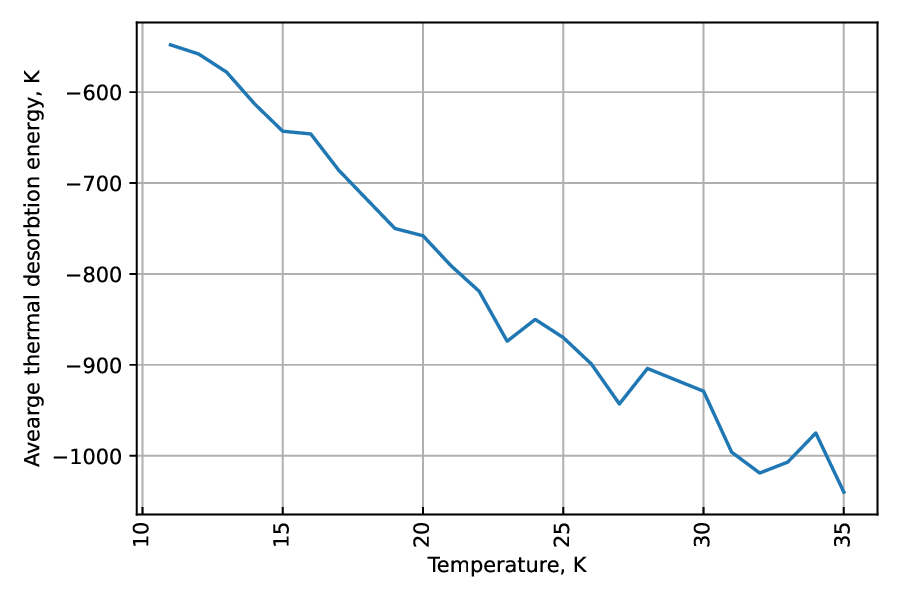} & \includegraphics[height=3.5cm,keepaspectratio]{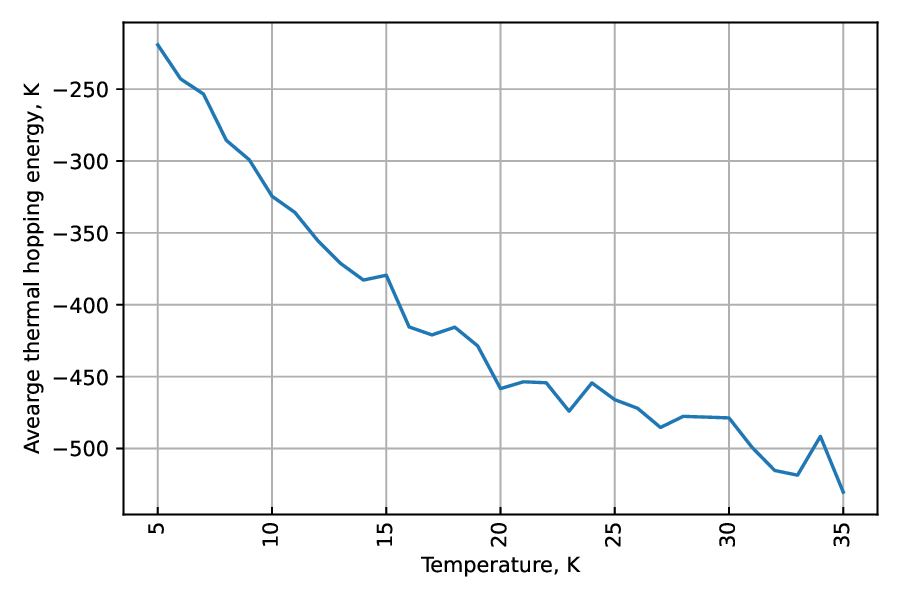} & \includegraphics[height=3.5cm,keepaspectratio]{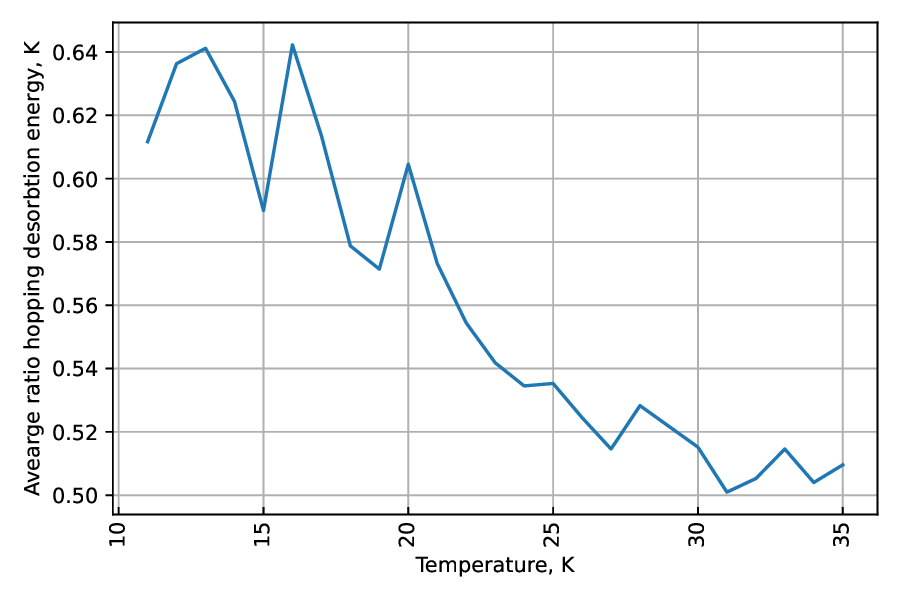}\\
(a)&(b)&(c)\\ 
\end{tabular}
    \caption{Average desorption energies (a), thermal hopping energies (b) and their ratios (c) of hydrogen atoms vs. temperature. Average energies of happened desorption and hopping events rise with temperature because the higher the temperature, the higher potential barriers are more likely to be overcomed.}\label{fig:av_eb_ed}
\end{figure*}
\begin{figure}
\includegraphics[height=5.5cm,keepaspectratio]{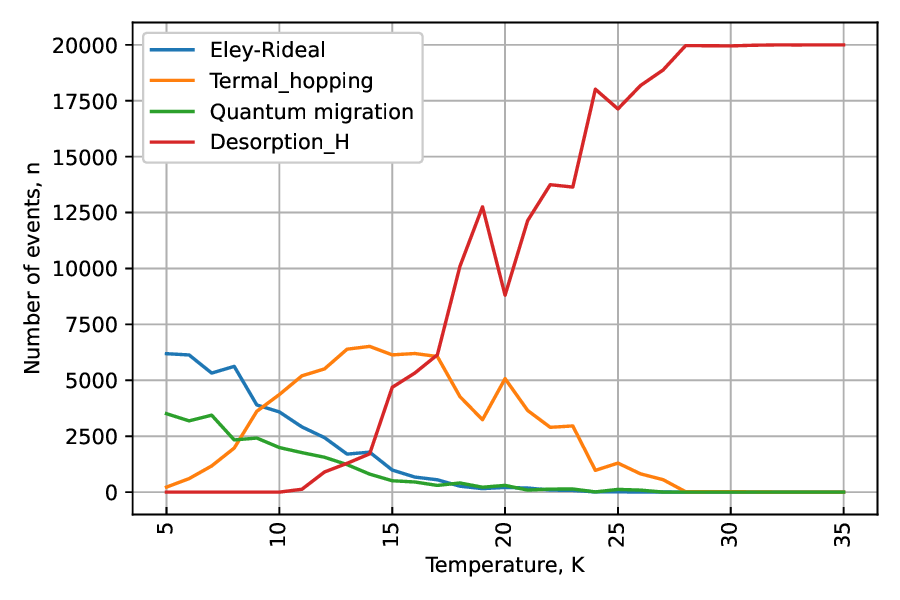}
\caption{Comparative efficiencies of desorption of atomic hydrogen and mechanisms of H$_{2}$ formation considered in the model vs. temperature. On vertical axis, a number of events that immediately led to the formation of H$_2$ molecules, and number of desorption events of atomic hydrogen are shown per 20\,0000 events of atomic hydrogen accretions from the gas phase.\label{fig:h2_form_mechanisms_vs_T}}
\end{figure}
\begin{figure}
\includegraphics[height=5.5cm,keepaspectratio]{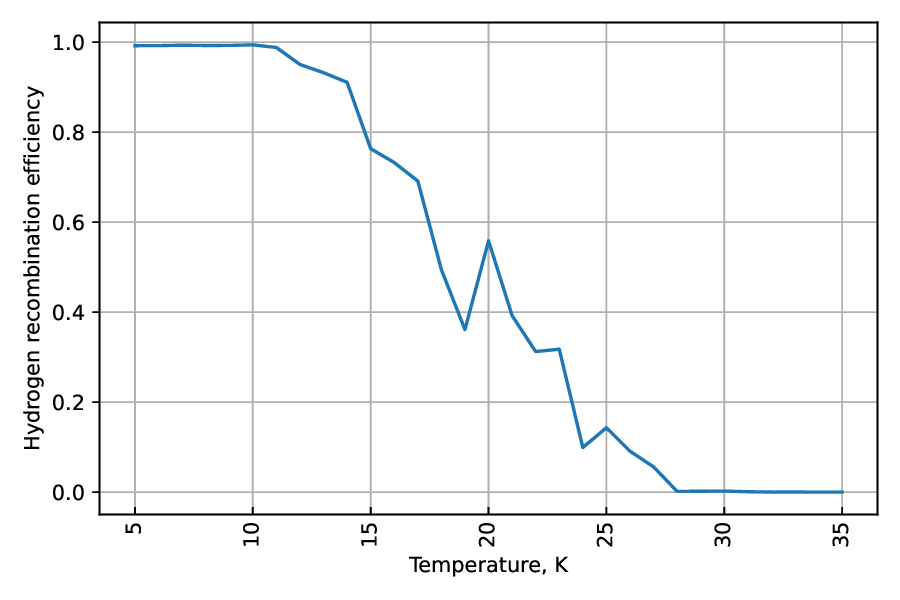}
\caption{H$_{2}$ formation efficiency vs. temperature for three gas densities.\label{fig:h2_form_Eff}}
\end{figure}
Since our model provides an information on energies of desorption and diffusion events, it is possible to calculate average energies for all happened events of desorption, as well as for all happened events of thermal hopping. Finally, average hopping-to-binding (diffusion-to-desorption) energy ratio can be obtained for every simulation. The averaged binding energies as well as hopping energies and their ratios vs. temperature obtained from our simulations are presented in Figure~\ref{fig:av_eb_ed}. It can be seen that average numbers of diffusion and desorption events get higher with temperature. It is reasonable, because on an uneven surface depths of potential minima are different, as previously shown at Figure~\ref{fig:distribution_of_potential_wells}. At higher temperatures, H atoms can desorb and diffuse out of deeper potential minima. The diffusion/desorption energy ratio stays relatively constant with temperature, although slightly decreasing at higher temperatures.

In Figure~\ref{fig:h2_form_mechanisms_vs_T}, the relative importance of different mechanisms of the formation of H$_2$ molecules is shown. It is estimated in a simplistic way. We attribute each particular event of H$_2$ formation to the process that immediately led to the encounter of two hydrogen atoms on grain surface. If an accreting hydrogen atom lands on top of already accreted hydrogen atom, the H$_2$ molecule is considered formed via Eley-Rideal mechanism. If an already accreted hydrogen atom tunnelled or thermally hopped from the currently occupied potential minimum to another potential minimum on surface that is sufficiently close to other hydrogen atom accreted on surface, the H$_2$ formation via diffusive Langmuir-Hinshelwood mechanism is considered to happen. Depending on the mechanism of final diffusion step -- tunnelling or thermal hopping -- H$_2$ formation is counted as ``via tunnelling'' or ``via thermal hopping''. By ``sufficiently close to other hydrogen atom'' we consider the situation when the value of interaction potential between two hydrogen atoms is larger than the value of total interaction potential between the surface and at least one of the two interacting hydrogen atoms.

It can be seen that at temperatures below 10~K, Eley-Rideal mechanism and diffusion of H atoms due to quantum tunnelling are the dominant mechanisms that lead to the formation of H$_2$ molecules. However, at higher temperatures, thermal diffusion of hydrogen atoms is the major mechanism that leads to the formation of molecular hydrogen. In Figure~\ref{fig:h2_form_Eff}, the overall H$_2$ formation efficiency vs. temperature is shown. The formation efficiency of molecular hydrogen is defined as twice the number of H$_2$ molecules divided by the number of accreted H atoms during the time span of simulation. Non-zero efficiency of H$_2$ formation is observed up to the dust temperature of 30~K. Obviously, this is possible due to the existence of deep potential minima on the uneven surface of our model grain. Note again, that only physisorption of hydrogen atoms is considered in our model.


\section{Discussion and Conclusions}

In this study, we presented an off-lattice microscopic Monte Carlo model of the formation of molecular hydrogen on a small bare carbonaceous interstellar dust grain. The key advantage of the off-lattice approach is the ability to simulate the interaction of hydrogen atoms with highly irregular amorphous surface. On such surface, energies and locations of binding sites are determined by mutual positions of adjacent atoms the surface consists of. Thus, binding energies of accreted hydrogen atoms vary at different binding sites. Similarly, energies that are needed to thermally diffuse from one binding site to another are different, too. Moreover, not only energies of binding sites are different, but also distances between them. Thus, diffusion due to quantum tunnelling through potential barriers between binding sites is efficient between the ``closely'' located sites, but inefficient if adjacent sites are located further away of each other.

Although our off-lattice Monte Carlo model is very simplistic in comparison to more sophisticated molecular dynamics models, we believe that it still can capture some key features of diffusion and desorption of hydrogen atoms on the irregular surface. The key findings of our simulations are the following. First, we observe a broad asymmetric distribution of energies of binding sites on the simulated grain surface. Minimum and maximum binding energies are differ by a factor of $\sim 3$. While the most probable binding energy value is closer to the minimum values, the broad ``tail'' of high binding energies exists. Such tail means that hydrogen atoms may stay at some sites on surface at temperatures significantly higher than a typical desorption temperature of hydrogen atoms. In turn, this makes possible the formation of H$_2$ molecules on grains at high temperatures. Therefore, roughness of surfaces of interstellar grains may play an important role at the formation of molecular hydrogen in warm astrophysical environments such as atomic and molecular diffuse clouds. This result is in line with previous studies by e.g. \citet[][]{2005MNRAS.361..565C, 2006A&A...458..497C, 2012ApJ...751...58I,2020A&A...634A..42T} who showed that the efficient formation of molecular hydrogen is possible over an extended range of grain temperatures if an uneven surface with strong binding sites is considered. Furthermore, a wide Gaussian-like distribution of binding energies observed in our model is consistent with recent detailed quantum chemical/molecular dynamics simulations by~\citet[][]{Groyne_ea25}, who obtained similar results for several molecular species on amorphous water ice. Earlier, \citet[][]{Tinacci_ea22}, obtained similar distributions for ammonia using the same modeling framework.

Second, the kinetics of diffusion of accreted hydrogen atoms in our model is not the same as in rate equations-based models. In RE models, surface is normally considered uniform, with equal binding energies at all binding sites. This means that a species (e.g. hydrogen atom) at certain temperature can diffuse across the surface on a long range and even infinitely ``scan'' the entire surface of a grain, until it meets another species on surface or desorb to gas. In our off-lattice model, we normally observe somewhat different picture of diffusion. As mentioned above, the simulated surface is irregular, and the binding sites on the surface may have significantly different energies. In our simulations, the most likely behaviour of diffusing hydrogen atoms is a relatively short-range diffusion that ends up in a high-energy binding site. Tunnelling through potential barriers that separate binding sites in our model is also of limited efficiency. This is also because of irregularity of the surface: not only depths of potential minima at binding sites are different, but also the thicknesses of potential barriers between them, too. As a result, thermal diffusion is a dominant mechanism of mobility for hydrogen atoms at grains with temperatures of 10~K and above. Interestingly, our conclusions on mobility of hydrogen atoms is consistent with the results of an elegant experimental study by \citet[][]{Kuwahata_ea15}. They found that although tunnelling contributes to the migration rate of hydrogen atoms, especially at short distances, the overall migration rate is mainly defined by thermal hopping. The limiting role of strong binding sites on the efficiency of migration of H atoms due to tunnelling is also highlighted in their study.

The results of our off-lattice Monte Carlo model, which highlight the importance of surface roughness in broadening the temperature range for efficient H$_2$ formation, align well with the findings of \citet[][]{Ivanovskaya_ea10} on topological defects in carbonaceous grains. Their work demonstrates that defects such as pentagonal rings and carbon adatoms significantly reduce energy barriers for H adsorption and H$_2$ recombination enabling catalysis even at cold interstellar temperatures. This supports our observation of a wide, asymmetric distribution of binding energies (520–-1800~K). While our model currently considers only physisorption, Ivanovskaya et al. show that chemisorption at defective sites could further enhance H$_2$ formation. Additionally, their proposed barrierless or low-barrier Eley-Rideal pathways complement our results, where both ER and Langmuir-Hinshelwood mechanisms contribute to H$_2$ production across a broad temperature range.

The possibility of H$_2$ formation from physisorbed hydrogen atoms on carbonaceous surfaces—via both the Eley-Rideal and Langmuir-Hinshelwood mechanisms, as proposed in our model—is not self-evident. However, DFT simulations by \citet{JeloaicaSidis99} confirmed that H$_2$ formation from physisorbed H atoms is indeed feasible, at least on graphite surfaces.

The model presented in this study is very simplistic. Real interatomic interactions cannot be fully described by the Lennard-Jones potentials. Usage of this potential, for example, makes impossible to include chemisorption in the current model. Moreover, the values of parameters $\sigma$ and $\epsilon$ that define the properties of the potential utilized in this study are chosen empirically. Therefore, the absolute values of binding energies obtained in this work shall not be considered as reference for other studies. However, other results of this study such as conclusion on a short-range nature of hydrogen diffusion on surface, prevalence of thermal hopping over quantum tunnelling as a mechanism of diffusion, and values of binding-to-desorption ratios have more general validity even though they obtained with the simplistic approach. Indeed, questions on whether or not quantum tunnelling for diffusion of H atoms shall be enabled in rate equations-based models, and which binding-to-desorption ratio shall be used are the long-standing questions in astrochemical modelling. The results of the off-lattice modelling presented in this study are aimed to contribute to the discussion.







\section*{Acknowledgments}
The authors thank Gleb Fedoseev and Juris Robert Kalnin for stimulating discussions. The authors also grateful to the anonymous reviewer for constructive comments that helped improve the manuscript. The work is supported by the Russian Science Foundation through Project 23-12-00315.

\section*{Data Availability}

The simulation data presented in this work are available upon request. 



\input{references.bbl}








\bsp	
\label{lastpage}
\end{document}